\documentstyle[12pt,epsf,cite]{article}

\newlength{\titlesep}
\newlength{\authorsep}
\makeatletter

\def\fnum@figure{FIG.~\thefigure}

\newcommand{\reffig}[1]{Fig.~\protect\ref{#1}}

\newcounter{figureparent}
\@addtoreset{figure}{figureparent}

\@addtoreset{equation}{section}

\newcounter{eqnparent}
\@addtoreset{equation}{eqnparent}

\renewcommand{\abstract}{\if@twocolumn
  \section*{Abstract}
  \else
  \begin{center}
    {\bf Abstract\vspace{-.5em}\vspace{0pt}}
  \end{center}
  \quotation
  \fi}
\renewcommand{\endabstract}{\if@twocolumn\else\endquotation\fi}

\newcommand{\thismonth}{\ifcase\month\or
 January\or February\or March\or April\or May\or June\or
 July\or August\or September\or October\or November\or December\fi
 \space \number\year}

\newcommand{\preprintnumber}[1]
{\begin{flushright}
  \begin{tabular}{l} #1 \end{tabular}
  \end{flushright}}

\makeatother

\newcommand{\Rn}[1]{{\uppercase\expandafter{\romannumeral#1}}}
\newcommand{\gsim}%
{\mathrel{\mbox{\raisebox{-1.0ex}%
{$\stackrel{\textstyle >}{\textstyle \sim}$}}}}
\newcommand{\lsim}%
{\mathrel{\mbox{\raisebox{-1.0ex}%
{$\stackrel{\textstyle <}{\textstyle \sim}$}}}}

\newcommand{\Journal}[4]{{#1} {\bf #2} {(#3)} {#4}}
\newcommand{\pl}{\sl Phys.~Lett.}
\newcommand{\plb}{\sl Phys.~Lett.~{\bf B}}

\newcommand{\pr}{\sl Phys.~Rev.}
\newcommand{\prd}{\sl Phys.~Rev.~{\bf D}}
\newcommand{\prl}{\sl Phys.~Rev.~Lett.}
\newcommand{\np}{\sl Nucl.~Phys.}

\newcommand{\ptp}{\sl Prog.~Theor.~Phys.}

\newcommand{\ep}{\sl Euro.~Phys.~J.~{\bf C}}

\newcommand{\pan}{\sl Phys.~Atom.~Nucl.}

\newcommand{\ibid}{\it ibid.}

\newcommand{\epsfile}[1]{\relax}

\begin{document}
\baselineskip 18pt

\begin{titlepage}
\preprintnumber{%
KEK-TH-608 \\
KEK-Preprint-98-206\\
}
\vspace*{\titlesep}
\begin{center}
{\LARGE\bf

Effect of supersymmetric CP phases on the
$B \to X_s \gamma$ and $B \to X_s l^+ l^-$ decays in the minimal
supergravity model
}
\\
\vspace*{\titlesep}
\footnote{E-mail: toru.goto@kek.jp.}Toru Goto,
\footnote{E-mail: keum@ccthmail.kek.jp, Monbusho fellow.}Y.-Y. Keum, 
\footnote{E-mail: niheit@ccthmail.kek.jp, JSPS fellow.}Takeshi Nihei,\\
\footnote{E-mail: yasuhiro.okada@kek.jp.}Yasuhiro  Okada, and
\footnote{E-mail: yasuhiro.shimizu@kek.jp.}Yasuhiro  Shimizu\\
\vspace*{\authorsep}
{\it Theory Group, KEK, Tsukuba, Ibaraki, 305-0801 Japan }
\\
\end{center}
\vspace*{\titlesep}
\begin{abstract}
We investigate the effect of supersymmetric CP violating phases on the
$B\to X_s \gamma$ and $B\to X_s l^+l^-$ decays in the minimal
supergravity model. 
We show that the phase of the trilinear scalar coupling constant for top
squarks is strongly suppressed and aligned to that of the gaugino mass
due to a renormalization effect from the Planck scale to the electroweak
scale. As a result, the effect of supersymmetric CP violating phases on
the $B \to X_s \gamma$ and  $B \to X_s l^+ l^-$ 
decays are small taking into account the neutron and the electron
electric dipole moment constraints. For the $B \to X_s\gamma$ decay, the
amplitude has 
almost no new CP violating phase and 
the direct CP asymmetry is less than 2 \%.
For the $B \to X_s l^+ l^-$ decay, the branching ratio can be 
sizably different from that in the standard model only when the sign of the
$B \to X_s \gamma$ amplitude is opposite to that in the standard model.

\end{abstract}
\end{titlepage}
%%%%%%%%%%%%%%%%%%%%%%%%%%%%%%%%%%%%%%%%%%%%%%%%%%%%%%%%%%%%%%%%%%%%%%
The origin of the CP violation is one of main issues in current particle
physics. In the standard model (SM) the CP violation is originated from the
phase of the Kobayashi-Maskawa matrix \cite{KM}. A new source of CP
violation can appear in models beyond the SM. 

Among various models beyond the SM, the minimal supersymmetric SM (MSSM)
is one of the most promising candidate. 
The MSSM contains many new parameters, {\it i.e.} soft supersymmetry
(SUSY) breaking parameters.
If we allow arbitrary soft SUSY breaking parameters, too large flavor
changing neutral current (FCNC) processes, such as $K^0\overline{K}^0$
mixing,  appear. 
In the minimal supergravity (mSUGRA), soft SUSY breaking parameters are
assumed to have universal structures at the Planck scale,
so that these dangerous FCNC processes are suppressed.

In the mSUGRA, there is no intrinsic
reason that these SUSY breaking parameters should be real and 
there are four new CP violating phases, {\it i.e.} 
phases of the gaugino mass, the higgsino mass parameter, the SUSY
breaking Higgs boson mass, and the trilinear scalar coupling constant,
of which two combinations are physically independent.
These phases induce the neutron and electron electric dipole moments
(EDMs).
There are many works on the constraints of the EDMs in the MSSM
\cite{mssm} as well as in the mSUGRA \cite{sugra,IN,FO,TN}.
It is shown that in the mSUGRA, if we take a phase convention that the
trilinear scalar coupling constant and the higgsino mass parameter have 
phases, $\phi_A$ and $\phi_\mu$, respectively, the constraint on
$\phi_\mu$ is much stronger than that on $\phi_A$ \cite{IN,FO,TN}.

It is known that the $B \to X_s \gamma$ process gives strong constraint
on the SUSY model. 
In particular, in Ref. \cite{GOST}, rare $B$ decays, such as 
$B \to X_s \gamma$ and $B \to X_s l^+l^-$, are studied in the mSUGRA without 
new CP violating phases. 
For the $B\to X_s \gamma$ decay, the SUSY contributions interfere with 
the amplitude in the SM  either constructively or destructively, and the
amplitude can change its sign. It is also shown that the $B \to X_s l^+l^-$
branching ratio is enhanced  compared to the SM prediction if the sign
of the $B\to X_s \gamma$ amplitude is opposite to that in the SM.
It is interesting to investigate the effect of the CP violating phases
on  various $B$ decays. In Ref. \cite{TN} one of the authors (T.N.)
analyzed effect of the SUSY CP violating phase on 
$B^0\overline{B}^0$ mixing, and showed that the effect is small. 
Recently a possibility of large direct CP asymmetry in the 
$B \to X_s \gamma$ process  is studied in the MSSM \cite{KN,acp}, 
and MSSM with SUGRA-motivated SUSY breaking terms \cite{ACO}. 

In this letter we investigate the effect of the SUSY CP violating parameter
on rare $B$ decays, $B \to X_s \gamma$ and $B \to X_s l^+ l^-$, in the
mSUGRA. 
In our analysis we require the universality
of  SUSY breaking terms at GUT scale and explicitly solve the
renormalization group equations (RGEs) to
determine the masses and the mixings of SUSY particles and also
require the condition for the radiative electroweak symmetry breaking.
We confirm that $\phi_\mu$ is strongly constrained by the
neutron and electron EDM bounds whereas $\phi_A$ is almost
unconstrained. However we  
show that the phase of the $A$-term for top squarks is reduced due to the
large top Yukawa coupling constant and aligned to that of
the gaugino mass. Therefore the phase of the $A$-term for top squarks is
strongly suppressed. We show that the CP asymmetry in rare $B$ decays
is suppressed in the
mSUGRA if the neutron and electron EDM constraints  are 
taken into account. 

In the MSSM, the Yukawa coupling constants are described by the
following superpotential, 
\begin{eqnarray}
  W_{\rm{MSSM}} = (Y_U)_{ij} Q_i U_j H_2 + (Y_D)_{ij} Q_i D_j H_1 +
  (Y_E)_{ij}E_i L_j H_1 - \mu H_1 H_2,
\end{eqnarray}
where $Q$ and $L$ denote the $SU(2)_L$ quark and lepton doublets,
$U$, $D$, and $E$ are $SU(2)_L$ singlets, 
and  $H_1$, $H_2$ are $SU(2)_L$ Higgs doublets.
The $i,j$ represent generation indices. In addition to the
SUSY invariant terms, there are following soft SUSY breaking terms,
\begin{eqnarray}
    -{\cal L}_{\rm soft} &=&
    (m_Q^2)_{ij} {\tilde q}^\dagger_{Li} {\tilde q}_{Lj}
  + (m_U^2)_{ij} {\tilde u}_{Ri}^{\ast} {\tilde u}_{Rj}
  + (m_D^2)_{ij} {\tilde d}_{Ri}^{\ast} {\tilde d}_{Rj}
\nonumber \\ 
  &+& (m_L^2)_{ij} ~{\tilde \ell}^\dagger_{Li} ~{\tilde \ell}_{Lj}
  + (m_E^2)_{ij} ~{\tilde e}_{Ri}^{\ast} ~{\tilde e}_{Rj}
\nonumber \\ 
  &+& \Delta^2_1  h^\dagger_1 h_1 
  + \Delta^2_2  h^\dagger_2 h_2
  + ( B \mu h_1 h_2 + {\rm H.c.} )
\nonumber \\ 
  &+& 
    \left[(A_{U})_{ij} {\tilde q}_{Li}  h_2\, {\tilde u}^\ast_{Rj}
  +  (A_{D})_{ij} h_1 {\tilde q}_{Li} {\tilde d}^\ast_{Rj}
  +  (A_{E})_{ij}  h_1 {\tilde e}^\ast_{Ri}{\tilde \ell}_{Lj}  +
  {\rm H.c.}\right]
\nonumber \\ 
  &+& (\frac{1}{2}M_1{\tilde B}{\tilde B}
  +  \frac{1}{2} M_2{\tilde W}{\tilde W}
  +  \frac{1}{2} M_3{\tilde G}{\tilde G} + {\rm H.c.}).
\end{eqnarray}
Hereafter we denote the superpartners by small letters with tilde.
The SUSY breaking terms depend on details of SUSY breaking mechanism.
In the minimal supergravity model these SUSY breaking terms are
originated from the gravitational interaction and given by the following 
universal structure at the high energy scale:
\begin{eqnarray}
  && M_1 = M_2 = M_3 = M_X, \\
  &&(m_Q^2)_{ij}=(m_U^2)_{ij}=(m_D^2)_{ij}=(m_L^2)_{ij}
  =(m_E^2)_{ij}=m_0^2\delta_{ij},\\  
  &&~~~\Delta_1^2 = \Delta_2^2 = m_0^2,\\
  &&(A_U)_{ij}=A_X(Y_U)_{ij}, ~(A_D)_{ij}=A_X (Y_D)_{ij},~
  (A_E)_{ij}=A_X (Y_E)_{ij}.
\end{eqnarray}
For simplicity we assume the GUT relation for gaugino masses and put the
universal condition at the GUT scale ($\simeq 2 \times 10^{16}$) neglecting
the renormalization effect from the Planck scale to the GUT scale. 
The SUSY breaking parameters at the electroweak scale are obtained by solving 
the RGEs.
In principle, the parameters, $M_X$, $A_X$, $\mu$, and $B\mu$, can have
 phases. Since only two combinations of the four phases are
physical CP violating phases, we take only $A_X$ and $\mu$ as complex
parameters hereafter.

In order to see qualitative feature of RGEs for $A$-terms let us first
neglect flavor mixings in the RGEs. The RGEs of $A$-terms for the first
and second generations are given by
\begin{eqnarray}
  \frac{d}{dt}A_{e_i} &=& 3 \alpha_2 M_2 + 3 \alpha_1 M_1 - 
  \alpha_{\tau} A_{\tau} -3 \alpha_{b} A_{b},
\\
\frac{d}{dt}A_{d_i} &=& \frac{16}{3} \alpha_3 M_3 + 3 \alpha_2 M_2
 + \frac{7}{9} \alpha_1 M_1 - \alpha_{\tau} A_{\tau} -3 \alpha_{b}A_b,
\\
  \frac{d}{dt}A_{u_i} &=& \frac{16}{3} \alpha_3 M_3 + 3 \alpha_2 M_2
 + \frac{13}{9} \alpha_1 M_1   - 3\alpha_{t} A_{t},
\end{eqnarray}
where $i=1,2$ and for the third generation
\begin{eqnarray}
  \frac{d}{dt}A_{\tau} &=& 3 \alpha_2 M_2 + 3 \alpha_1 M_1 - 4
  \alpha_{\tau} A_{\tau} -3 \alpha_{b} A_{b},
\\
\frac{d}{dt}A_{b} &=& \frac{16}{3} \alpha_3 M_3 + 3 \alpha_2 M_2
 + \frac{7}{9} \alpha_1 M_1 - \alpha_{\tau} A_{\tau} -6 \alpha_{b}
 A_{b} - \alpha_{t} A_{t},
\\
  \frac{d}{dt}A_{t} &=& \frac{16}{3} \alpha_3 M_3 + 3 \alpha_2 M_2
 + \frac{13}{9} \alpha_1 M_1  -\alpha_{b} A_{b} - 6\alpha_{t} A_{t}.
\end{eqnarray}
Here $A_{f_i}\equiv (A_{f})_{ii}/(Y_{f})_{ii}$, $t=-\ln (Q^2)/(4\pi)$
where $Q$ is a renormalization point, $\alpha_i=g_i^2/(4\pi)$, and
$\alpha_{f_i}= Y^2_{f_{ii}}/(4\pi)$. In the right-hand sides (RHS's) of
above equations, only the Yukawa coupling constants of the third
generation are retained. Since the RHS's of RGEs for $A$-terms depend
linearly on the $A$-terms and gaugino masses, general solutions can be
written in terms of the universal $A$-term and gaugino mass as follows: 
\begin{eqnarray}
  \label{aterm}
  A_{f_i} = C^A_{f_i} A_X - C^g_{f_i} M_X,
\end{eqnarray}
where the coefficients $C^A_{f_i}$ and $C^g_{f_i}$ are functions of the
Yukawa and gauge coupling constants. In Fig.~\ref{fig:cgcA-tanb},
$C^g_{f_i}$ and $C^A_{f_i}$ are shown as a function of $\tan\beta\
(=v_2/v_1)$.  
$C^A_t$ is much smaller than  $C^g$ because $C^A_t$ is reduced by the
large top Yukawa coupling constant. 
Therefore, the phase of $A$-term for top squarks is strongly
suppressed due to the renormalization effects even if the phase of
$A$-term at $M_X$ scale is maximal. Considering the current experimental
lower bound on the chargino mass, $m_{{\tilde \chi^+}} \gsim 91$ GeV
\cite{OPAL}, $M_X$ at the GUT scale must be roughly larger than 120
GeV. In principle, the contribution from $A_X$ can dominate in
Eq. (\ref{aterm}) if $A_X$ is larger than $O(10)$ TeV.
However, it makes scalar particles heavier than 1 TeV,
in which case SUSY loop effects on FCNC processes in $B$ decays are small. 
As $\tan\beta$ becomes larger, $C^A_{f_i}$ for the bottom squark is
reduced  due to the bottom Yukawa coupling constant. 
Note that the suppression of the phase of $A_t$ is general feature in
models where the $A$-terms are generated at a high energy scale.

Let us discuss phenomenological consequences of SUSY CP phases on 
the $B\to X_s \gamma$ and $B \to X_s l^+ l^-$ decays. 
These processes are described by the following effective Lagrangian
\begin{eqnarray}
  {\cal L}_{eff} = \frac{4G_F}{\sqrt{2}} V_{tb}V_{ts}^*\sum_{i=1}^{10}
  C_i(Q) O_i(Q).
\end{eqnarray}
The operators, $O_7$-$O_{10}$, are most relevant for the calculation
of the processes, which are given by,
\begin{eqnarray}
  O_7&=&\frac{e}{16\pi^2}m_b \overline{s_L} \sigma^{\mu\nu}b_R
  F_{\mu\nu},
\\
  O_8&=&\frac{g_3}{16\pi^2}m_b \overline{s_L} \sigma^{\mu\nu}T^a b_R
  F^a_{\mu\nu},
\\
  O_9&=&\frac{e^2}{16\pi^2} \overline{s_L}\gamma_\mu b_L
  \overline{l}\gamma^\mu l,
\\
  O_{10}&=&\frac{e^2}{16\pi^2} \overline{s_L}\gamma_\mu b_L
  \overline{l}\gamma^\mu\gamma_5 l.
\end{eqnarray}
In order to calculate the $B\to X_s \gamma$ and  $B\to X_s l^+l^-$
processes in the mSUGRA, we first solve the RGEs of the MSSM to
determine the masses and mixings of SUSY particles. Then, integrating
out SUSY particles at the electroweak scale, the SUSY contributions are
included into the Wilson coefficients $C_i$ in matching conditions.
The Wilson coefficients at the bottom mass scale are
calculated by solving the RGE of QCD at the next-to-leading order (NLO). 
As for the NLO calculation we follow the results in Ref. \cite{KN,CMM}
for the $B\to X_s \gamma$ process, and the results in Ref. \cite{bsll}
for $B\to X_s l^+l^-$ process. 

The direct CP asymmetry in the $B \to X_s \gamma$ decay is given by
\cite{KN}
\begin{eqnarray}
  \label{acp}
  A_{CP}(\delta) &=& \frac{\Gamma (\overline{B} \to X_s  \gamma) 
    - \Gamma (B \to X_{\overline{s}}\gamma)}
  {\Gamma (\overline{B} \to X_s \gamma)
    + \Gamma (B \to X_{\overline{s}}  \gamma)},
\nonumber\\
  &=&
  \frac{\alpha_3(\mu_b)}{|C_7|^2} 
  \left[\frac{40}{81}\mbox{Im}(C_2 C_7^*) -
    \frac{8z}{9}\left[v(z)+b(z,\delta)\right] \mbox{Im}\left[(1+\epsilon_s)C_2 C_7^*\right]
\right.
\nonumber
\\&&
\left.
   -\frac{4}{9}\mbox{Im}(C_8 C_7^*) + \frac{8z}{27}b(z,\delta)
  \mbox{Im}\left[(1+\epsilon_s)C_2 C_8^*\right]
\right],
\end{eqnarray}
where $\delta$ is an energy cutoff parameter for the photon,
$\mu_b$ is a renormalization point at the bottom mass scale,
$z=(m_c/m_b)^2$, $\epsilon_s = V_{ub}V_{us}^*/(V_{tb}V_{ts}^*)$,
and functions $v$ and $b$ are found in Ref. \cite{KN}.
In the SM, the CP asymmetry is estimated as $A_{CP}^{SM}(\delta=0.99)
\simeq 1.5 \times 10^{-2} \eta$ where $\eta$ is the Wolfenstein parameter.
The SM prediction is small because the small parameter $\epsilon_s$, which
is $O(10^{-2})$, is the only source of the direct CP violation in the $B
\to X_s \gamma$ process. If $C_7$ or $C_8$ has a sizable new CP
violating phase, $A_{CP}$ could be large. 

The dilepton spectrum of the $B \to X_s l^+ l^-$ decay can be written by
\begin{eqnarray}
  \label{decayrate}
&& \frac{d\, {\bf B} (B \to X_s  l^+ l^-)}{d\, \hat{s}}
=  {\bf B} (B \to X_c l  \overline{\nu})
        \frac{\alpha^2}{4 \pi^2}\left|\frac{V_{tb}V_{ts}^*}{V_{cb}}\right|^2
         \frac{1}{f(m_c/m_b)\kappa(m_c/m_b)}(1-\hat{s})^2
\nonumber\\ 
  &&~~~~~~ \times \left[ \vphantom{\frac{1}{1}}
         (|C_9^{eff}|^2 + |C_{10}|^2) (1+2 \hat{s})
         + \frac{4}{\hat{s}} |C_7|^2 (2+ \hat{s})
        + 12 \mbox{Re}(C_7^\ast C_9^{eff}) \right], 
\end{eqnarray}
where $\hat{s}$ is the dilepton invariant mass square normalized by
bottom mass square, $f=1-8x^2+8x^6-x^8-24x^4\ln x$, and 
$\kappa$ is a QCD correction factor \cite{bsll}. Since there is an
interference term between $C_7$ and $C_9^{eff}$, the dilepton spectrum
depends on the phases of  $C_7$ or $C_9^{eff}$.

For numerical calculations in the mSUGRA, we scanned the SUSY parameters
in the range $0<m_0<1$ TeV, $0< M_X <0.5$ TeV, $|A_X|<5 m_0$,
and we follow Ref. \cite{GOST} for detailed procedures of the calculation.
For definiteness, we take the KM parameters, $|V_{us}|=0.2196$,
$|V_{cb}|=0.0395$, $|V_{ub}/V_{cb}|=0.08$, and $\delta_{13}=\pi/2$ in the
standard parametrization \cite{PDG}. As for the $B\to X_s\gamma$ decay,
we take $\delta=0.99$, $\mu_b=m_b=4.8$ GeV, and $m_t$ = 175 GeV.

In Fig. \ref{fig:dn-phAmu} the neutron EDM is shown as a function of
$\phi_\mu$ and $\phi_A$ for $\tan\beta=30$. In the numerical calculation
only the neutron EDM from the quark EDMs is included. 
Recently it was pointed out that the EDM constraints may be relaxed by a 
cancellation among different contributions in the mSUGRA \cite{IN,FO}.
However, we do not rely on such a
cancellation because each contribution has different hadronic
uncertainty,
so that it is difficult to determine the parameters where such a
cancellation occurs. We are only interested in the case
where SUSY particles are lighter than 1 TeV
because otherwise SUSY effects on the $B\to X_s \gamma$ and $B \to X_s
l^+ l^-$ decays are strongly suppressed.
In such a case, as
pointed out in Ref. \cite{IN,FO,TN}, the neutron EDM exceeds the present
experimental bound, $|d_n| \le 0.97 \times 10^{-25} \ e$ cm \cite{dn}, unless
$\phi_\mu$ is $\lsim 10^{-2}$. On the other hand, $\phi_A$ can be $O(1)$
even if the masses of SUSY particles are $O(100)$ GeV.

The real and imaginary parts of $C_7$ at the bottom 
mass scale divided by the SM value are plotted in Fig. \ref{fig:complexC7}
for $\tan\beta$=3, 10, 30.
As in the case of no SUSY CP violating phase, 
the SUSY contributions to $C_7$ and $C_8$ become large, however, those
to $C_9$ and $C_{10}$ are small. In this figure,
the experimental bound on the $B\to X_s \gamma$ branching ratio,
$2.0 \times 10^{-4}$ $<$ ${\bf B}(B\to X_s\gamma)$ $<$ $4.5 \times 10^{-4}$
\cite{CLEO}, is imposed, therefore the region between
two circles are allowed. It is interesting to see that without the
neutron and electron EDM constraints, $C_7$ can have different 
phase from the SM value. On the other hand, 
with the EDM constraints, {\it i.e.}, $|d_n|<0.97 \times 10^{-25}\ e$ cm, 
$|d_e| < 4.0 \times 10^{-27}\ e$ cm \cite{de},
the imaginary part of $C_7/C_7^{SM}$ 
is quite suppressed, and either $C_7 \simeq C_7^{SM}$ or $C_7 \simeq
-C_7^{SM}$ region is allowed. This is a similar result
to that obtained without the SUSY CP violating phases \cite{GOST}. 
It is known that the charged Higgs boson and the chargino contributions
to $C_7$ can be significant, and that the charged Higgs boson
contribution to $C_7$ has the same phase as the SM contribution.
On the other hand the chargino-stop loop contribution to $C_7$ depends
on the new SUSY CP phases. 
In order to have the large phase of $C_7$,
the imaginary part of the chargino contribution must be large.
With the neutron and electron EDM constraints,
however, $\phi_\mu$ must be quite small. Moreover the phase of $A_t$ 
is also suppressed due to the RGE effect as discussed above.
Therefore it is difficult to have large phase of $C_7$.
We also find that $C_8/C_8^{SM}$ does not induce large imaginary part
although the magnitude itself can be changed by the SUSY contributions.
This means that large CP 
violating phases do not appear in rare $B$ decays in the framework of
the mSUGRA even if the new CP violating phases are introduced. This is a
distinct feature from the result which is obtained in Ref. \cite{ACO}
where the authors did not follow the universality condition for scalar
masses at the GUT scale.
From Fig. 3(a)-(c), only for $\tan\beta$ =30, there is parameter space
where $C_7/C_7^{SM}$ is negative.
We find that in this parameter region the lighter stop mass is less than
about 200 GeV as in the case of no new SUSY CP phase \cite{GOST}.

In Fig. \ref{fig:bsllAcp-dn}(a)-(b), we plot the 
$B\to X_s \mu^+ \mu^-$ branching ratio and the direct CP asymmetry in
the $B \to X_s \gamma$ versus the neutron EDM for $\tan\beta=30$.
After taking into account the neutron and electron EDM constraints,
we show that there are two branches of the $B \to X_s \mu^+ \mu^-$
branching ratio and the larger branching ratio corresponds to the case
where the sign of $C_7$ is opposite to that in the SM.
The branching ratio can be about twice as large as that in the SM
with $C_7 \simeq - C_7^{SM}$. 
The CP asymmetry turns out to be less than 2 \%.

In conclusion, we investigate the effect of the SUSY CP violating phases 
($\phi_\mu, \phi_A)$ on the rare $B$ decays in the mSUGRA model
taking into account the RGEs for the SUSY breaking parameters.
If the SUSY particles are in the hundred GeV region, $\phi_\mu$ is
strongly constrained by the EDM bounds. On the other hand, 
the phase of $A$-term for top squarks is aligned to that of the gaugino
masses due to the RGEs. As a consequence, the effect of the SUSY CP
violating phases is small and either $C_7\simeq C_7^{SM}$ or 
$C_7\simeq -C_7^{SM}$ is allowed. We show that the direct  CP asymmetry 
is less than 2 \% taking into account the EDM constraints.
For the $B\to X_s l^+ l^-$ decay, there is a twofold ambiguity of the
branching ratio according to the sign of $C_7$. The branching ratio can
be twice as large as the SM value when $C_7\simeq -C_7^{SM}$.

The work of T.G. was supported in part by the Soryushi
Shogakukai. Y.Y.K. would like to thank M. Kobayashi for his hospitality
and encouragement. The works of T.N. and Y.Y.K. were supported in part by
the Grant-in-Aid of the Ministry of Education, Science, Sports and
Culture, Government of Japan. The work of Y.O. was supported in part by
the Grant-in-Aid of the 
Ministry of Education, Science, Sports and Culture, Government of Japan
(No.09640381), Priority area ``Supersymmetry and Unified Theory of
Elementary Particles'' (No. 707), and ``Physics of CP Violation''
(No.09246105).

\newpage

\section*{Figure Captions}

\newcounter{FIG}
\begin{list}{{\bf FIG. \arabic{FIG}}}{\usecounter{FIG}}
\item
The coefficients $C^A$ and $C^g$ in
      Eq. (\protect{\ref{aterm}}) are shown as a function of $\tan\beta
      $. Here  we take $\overline{MS}$ masses for quarks as $m_u=3.3$
      MeV, $m_d=6.0$ MeV, and $m_s=120$ MeV at the scale of 2 GeV. 
      We take pole masses as $m_c=1.4$ GeV and $m_b=4.8$ GeV, $m_t=175$ GeV.
\label{fig:cgcA-tanb}
\item
  The absolute value of the neutron EDM ($|d_n|$) is plotted as a function of
      $\phi_\mu$ (a) and  $\phi_A$ (b)
      for $\tan\beta=30$. Here input SUSY parameters are
      scanned in a region, $0<m_0<1$ TeV, $0<M_X<0.5$ TeV, and $|A_X|<5m_0$.
      The dashed line represents the present experimental upper bound,
      $|d_n|<0.97 \times 
      10^{-25}\ e$ cm. 
      For Fig. 2(b), squares correspond to the parameter spaces
      $\phi_\mu=0,\pi$.\label{fig:dn-phAmu}
\item
$C_7/C_7^{SM}$ at the bottom mass scale is shown imposing
       the current experimental bound for the $B \to X_s \gamma$
       branching ratio for $\tan\beta=3$ (a), 10 (b), 30 (c).
       Dots correspond to values without the
       neutron and electron EDM constraints and squares correspond to
       values with the EDM constraints. The input SUSY parameters are the
       same as  Fig. 2.\label{fig:complexC7}
\item
(a). The $B\to X_s \mu^+ \mu^-$ branching ratio is
  plotted as a 
      function of the neutron EDM. As for the $B\to X_s \mu^+ \mu^-$
      branching ratio, in order to avoid the $J/\psi$ resonance,
      we integrate the dilepton spectrum in a region $4 m_\mu^2 < s <
      (m_{J/\psi} - 0.1 (\mbox{GeV}))^2$ where $s$ is the dilepton invariant
      mass square. (b). The absolute value of the direct CP asymmetry in the
      $B\to X_s\gamma$ ($|A_{CP}|$)  
      is plotted as a function of the neutron EDM.
      In these figures the input SUSY parameters are the same as
      \reffig{fig:dn-phAmu}. The dashed line represents the present upper
      bound for the neutron EDM. Squares correspond to the parameter
      spaces with the electron and neutron EDM constraints.\label{fig:bsllAcp-dn}\\
\end{list}

%\clearpage
%\section*{Figures}
\pagestyle{empty}

\def\EPSDIR{}
\def\EPSSCALE{0.70}
\def\fnum@figure{FIG.~\thefigure}

~
\vfill
\begin{center}
\makebox[0cm]{
\def\epsfsize#1#2{\EPSSCALE#1}
\epsfbox{\EPSDIR 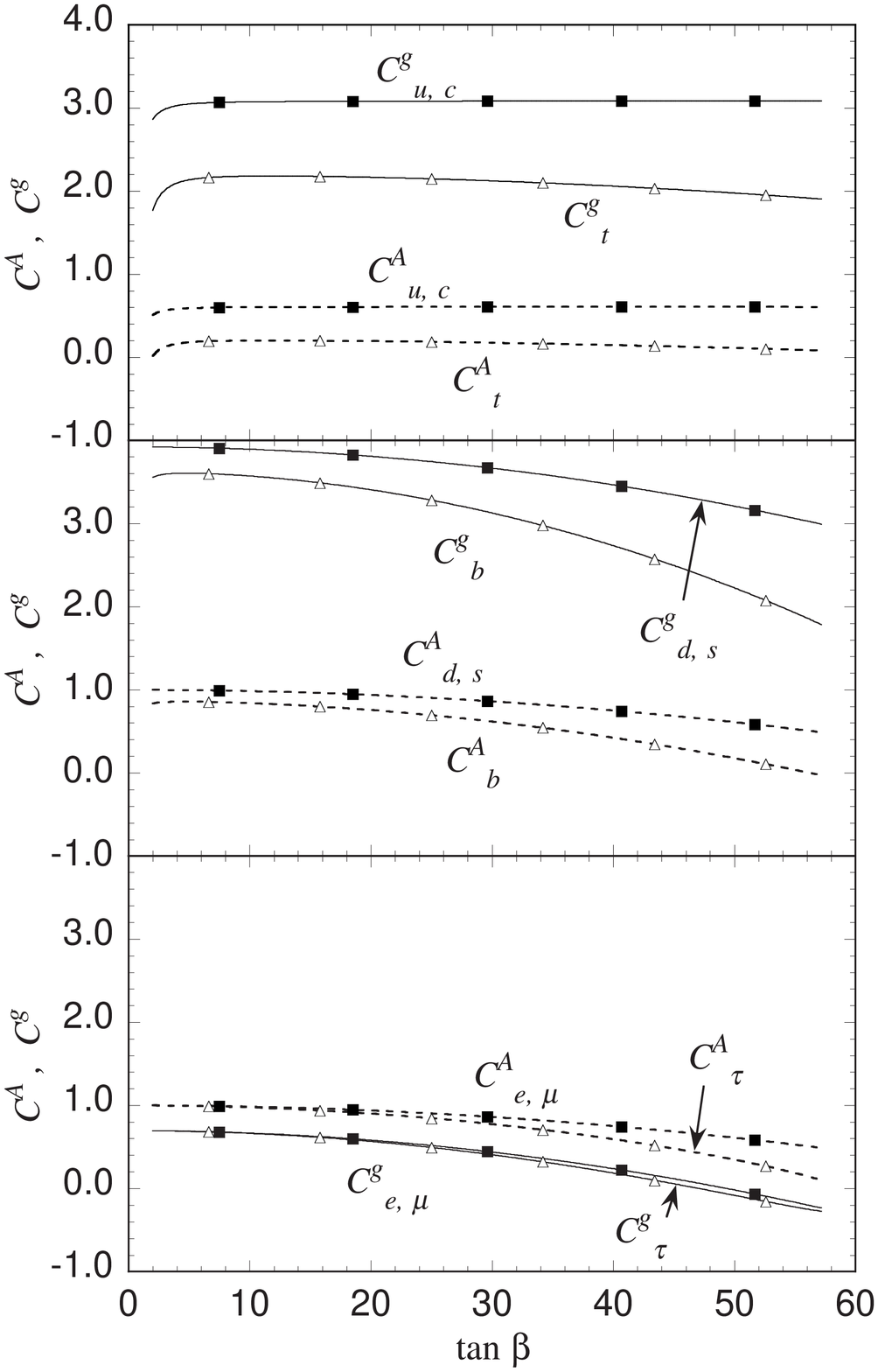}
}
\vfill
{\Large\bf Fig.~\ref{fig:cgcA-tanb}}
\end{center}
\clearpage

\def\EPSSCALE{1.0}

~
\vfill
\begin{center}
\makebox[0cm]{
\def\epsfsize#1#2{\EPSSCALE#1}
\epsfbox{\EPSDIR 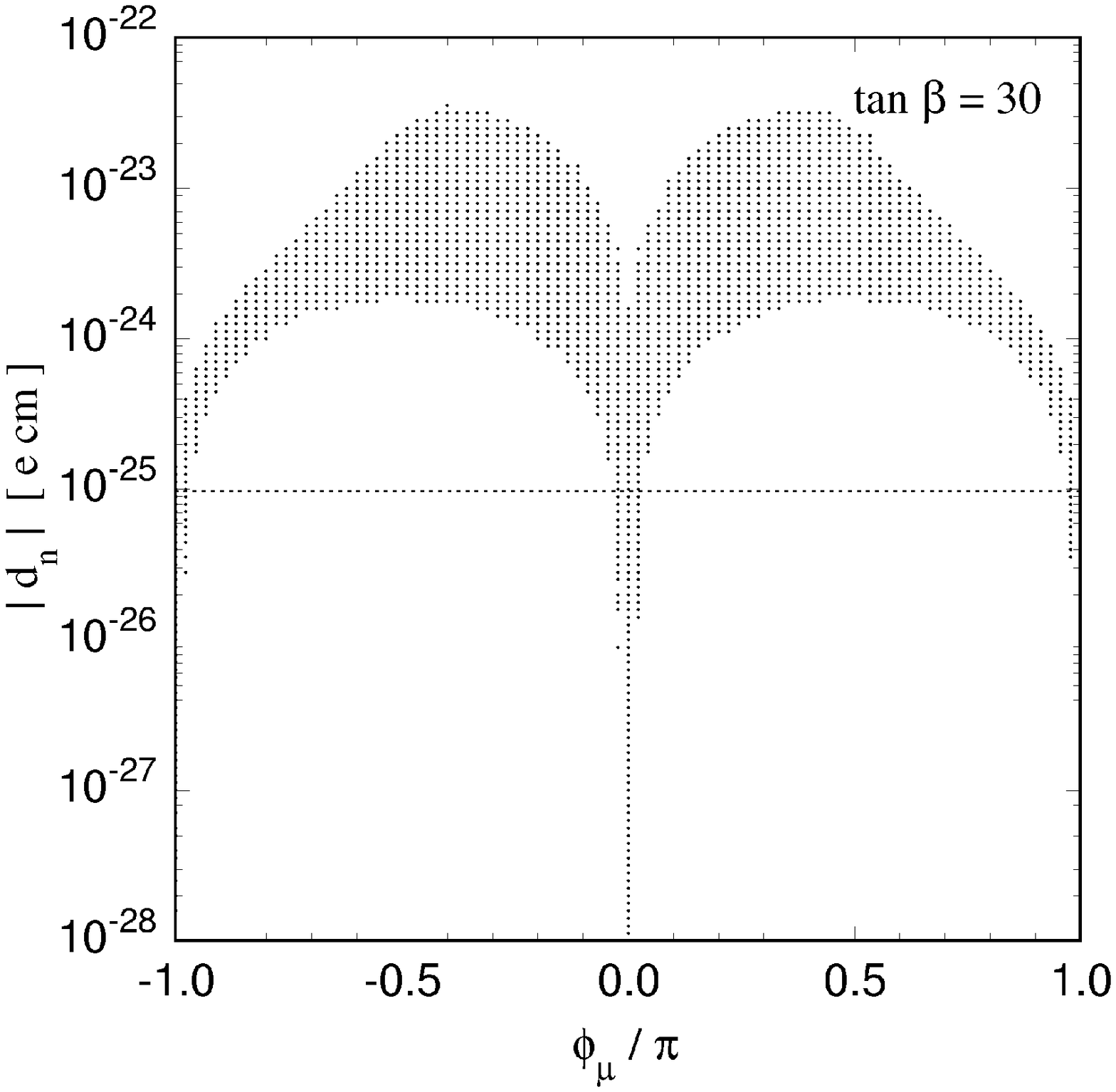}
}
\vfill
{\Large\bf Fig.~\ref{fig:dn-phAmu}(a)}
\end{center}
\clearpage

~
\vfill
\begin{center}
\makebox[0cm]{
\def\epsfsize#1#2{\EPSSCALE#1}
\epsfbox{\EPSDIR 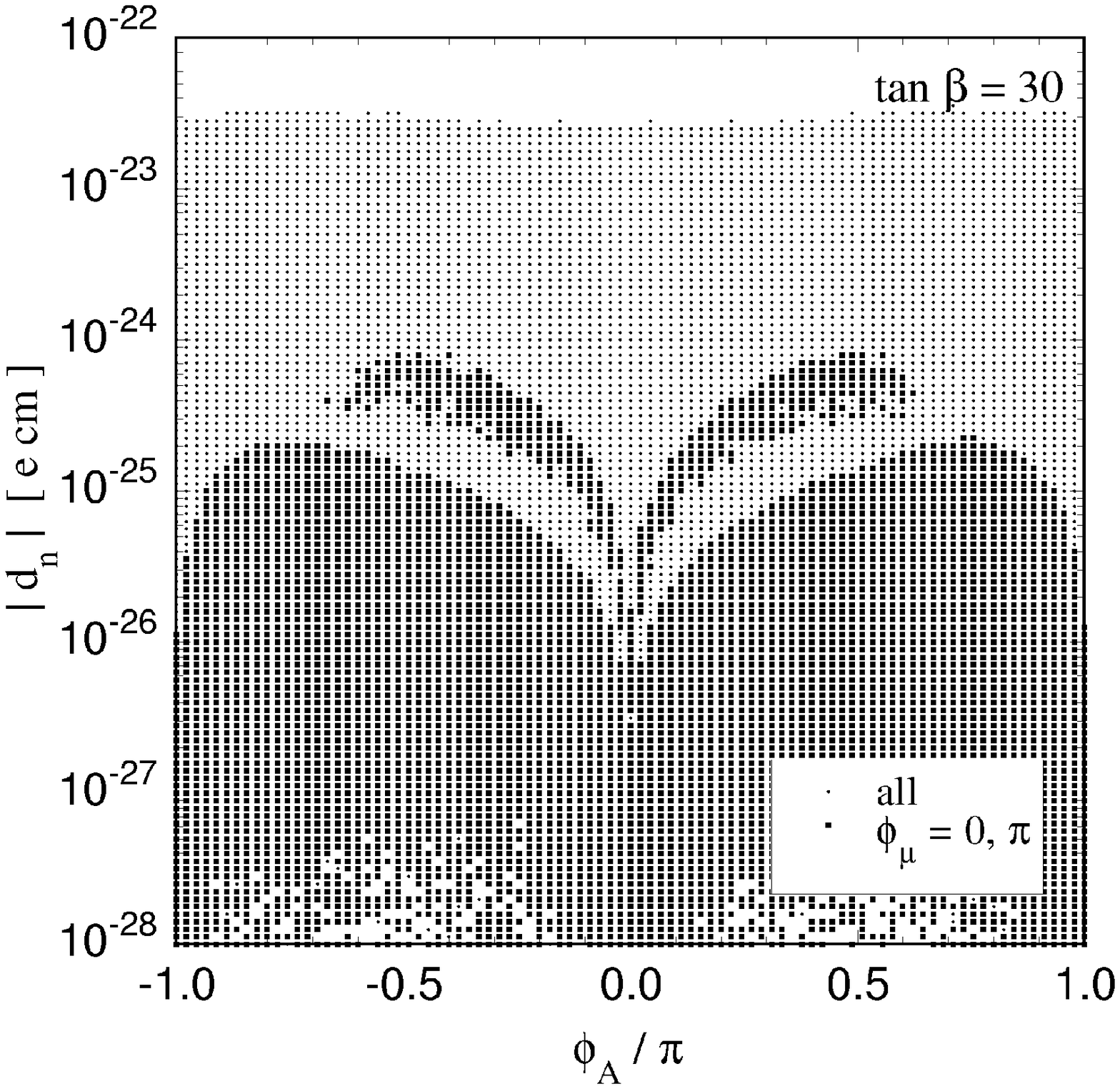}
}
\vfill
{\Large\bf Fig.~\ref{fig:dn-phAmu}(b)}
\end{center}
\clearpage

~
\vfill
\begin{center}
\makebox[0cm]{
\def\epsfsize#1#2{\EPSSCALE#1}
\epsfbox{\EPSDIR 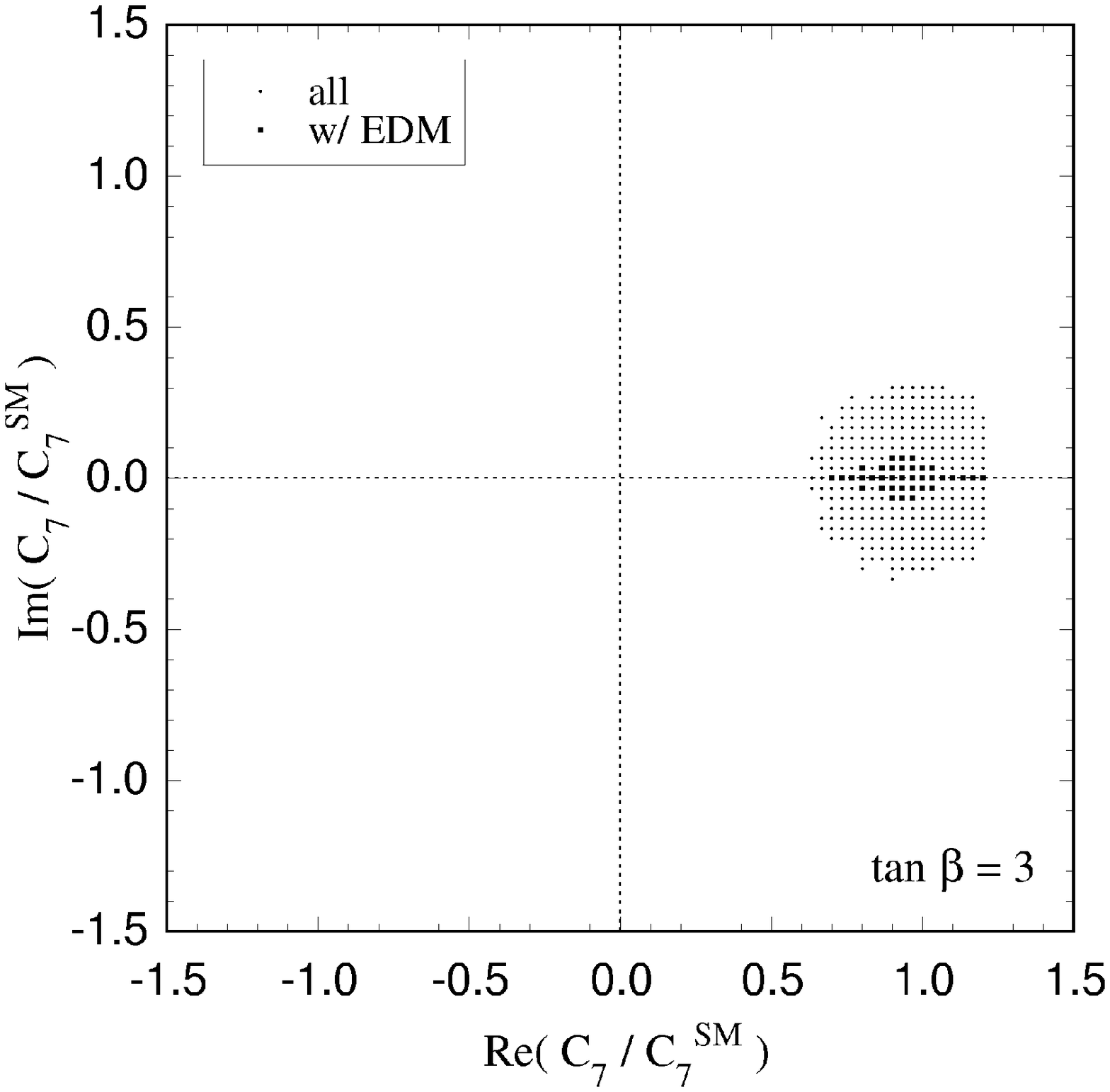}
}
\vfill
{\Large\bf Fig.~\ref{fig:complexC7}(a)}
\end{center}
\clearpage

~
\vfill
\begin{center}
\makebox[0cm]{
\def\epsfsize#1#2{\EPSSCALE#1}
\epsfbox{\EPSDIR 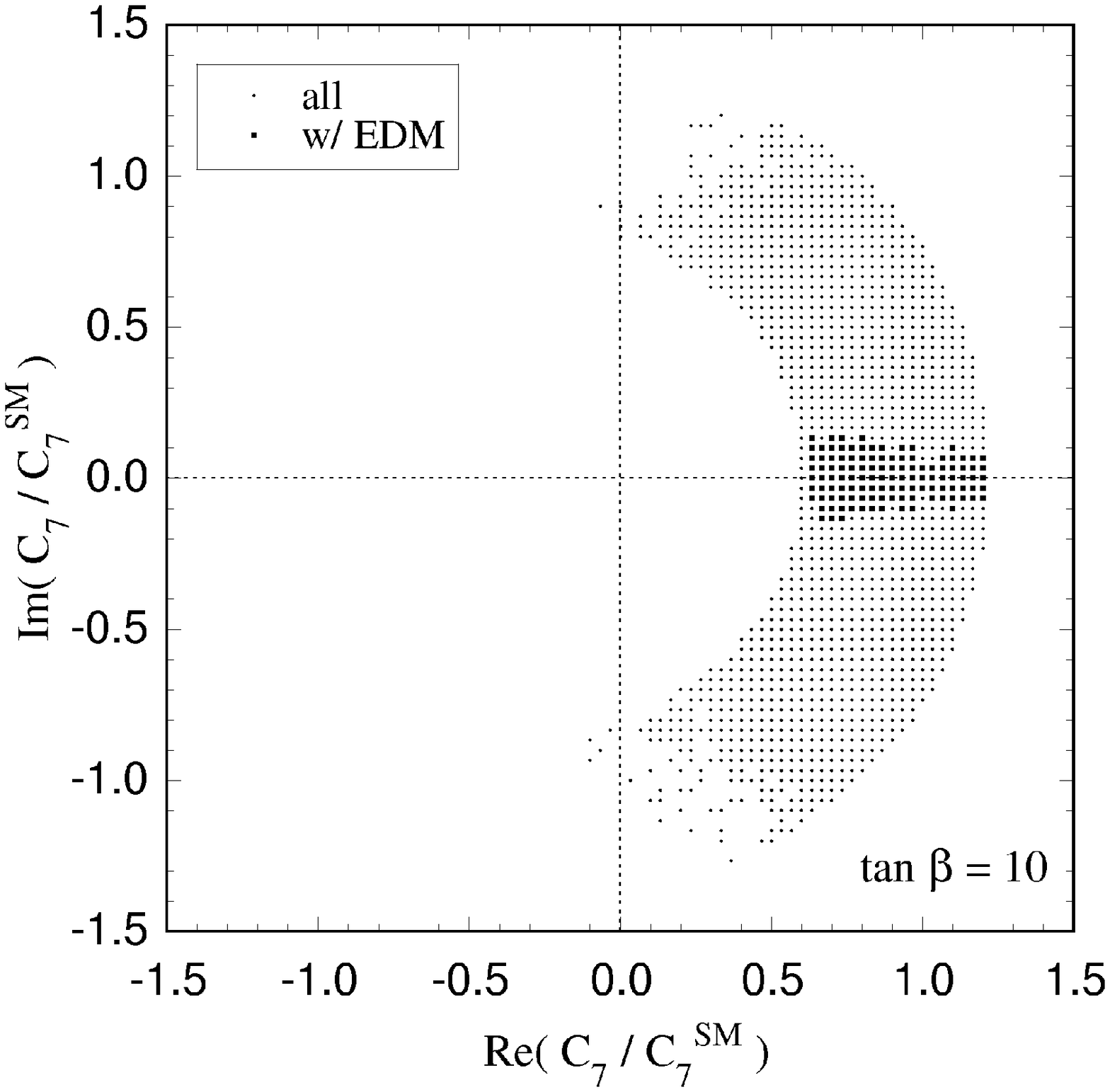}
}
\vfill
{\Large\bf Fig.~\ref{fig:complexC7}(b)}
\end{center}
\clearpage

~
\vfill
\begin{center}
\makebox[0cm]{
\def\epsfsize#1#2{\EPSSCALE#1}
\epsfbox{\EPSDIR 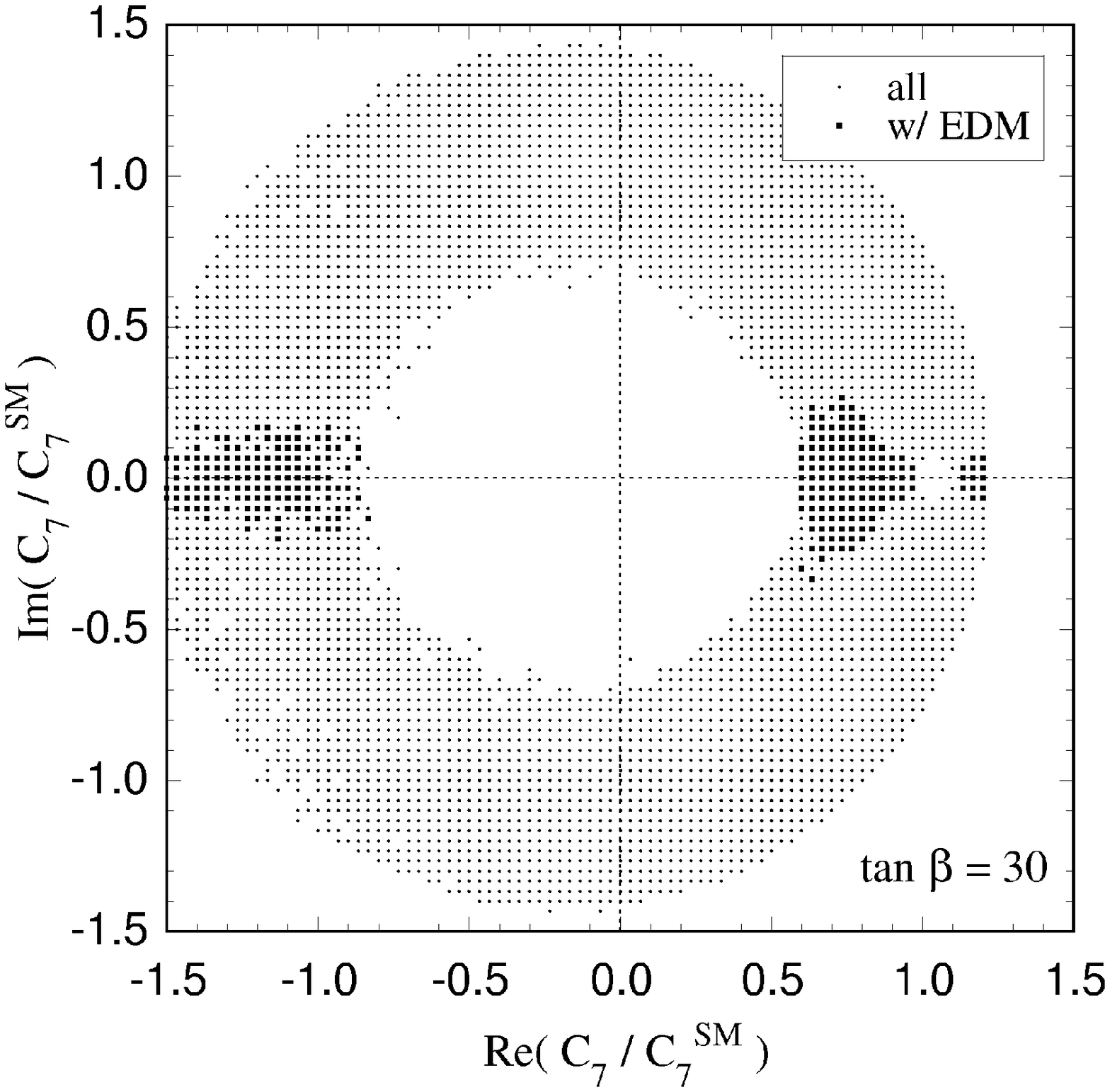}
}
\vfill
{\Large\bf Fig.~\ref{fig:complexC7}(c)}
\end{center}
\clearpage

~
\vfill
\begin{center}
\makebox[0cm]{
\def\epsfsize#1#2{\EPSSCALE#1}
\epsfbox{\EPSDIR 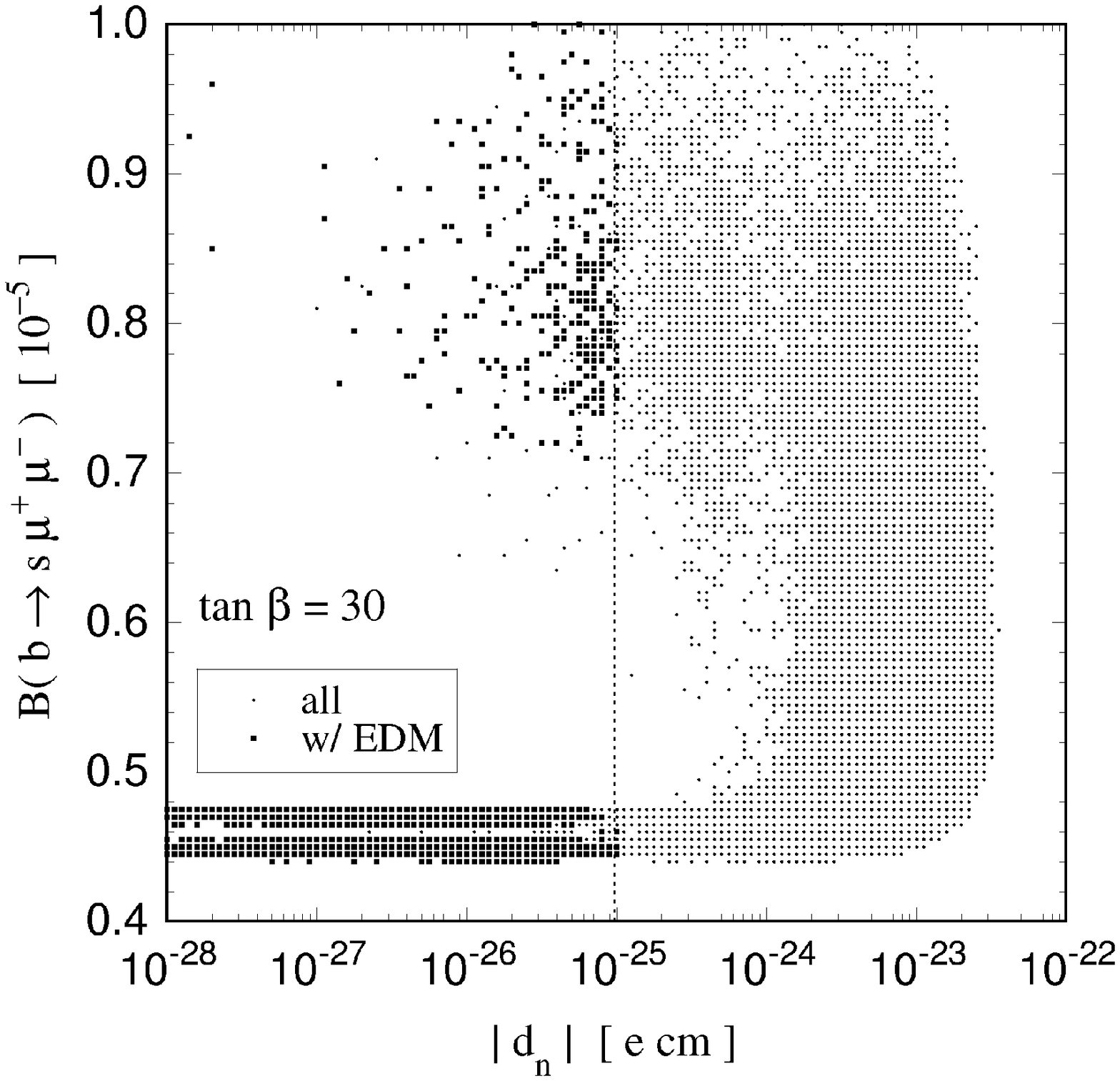}
}
\vfill
{\Large\bf Fig.~\ref{fig:bsllAcp-dn}(a)}
\end{center}
\clearpage

~
\vfill
\begin{center}
\makebox[0cm]{
\def\epsfsize#1#2{\EPSSCALE#1}
\epsfbox{\EPSDIR 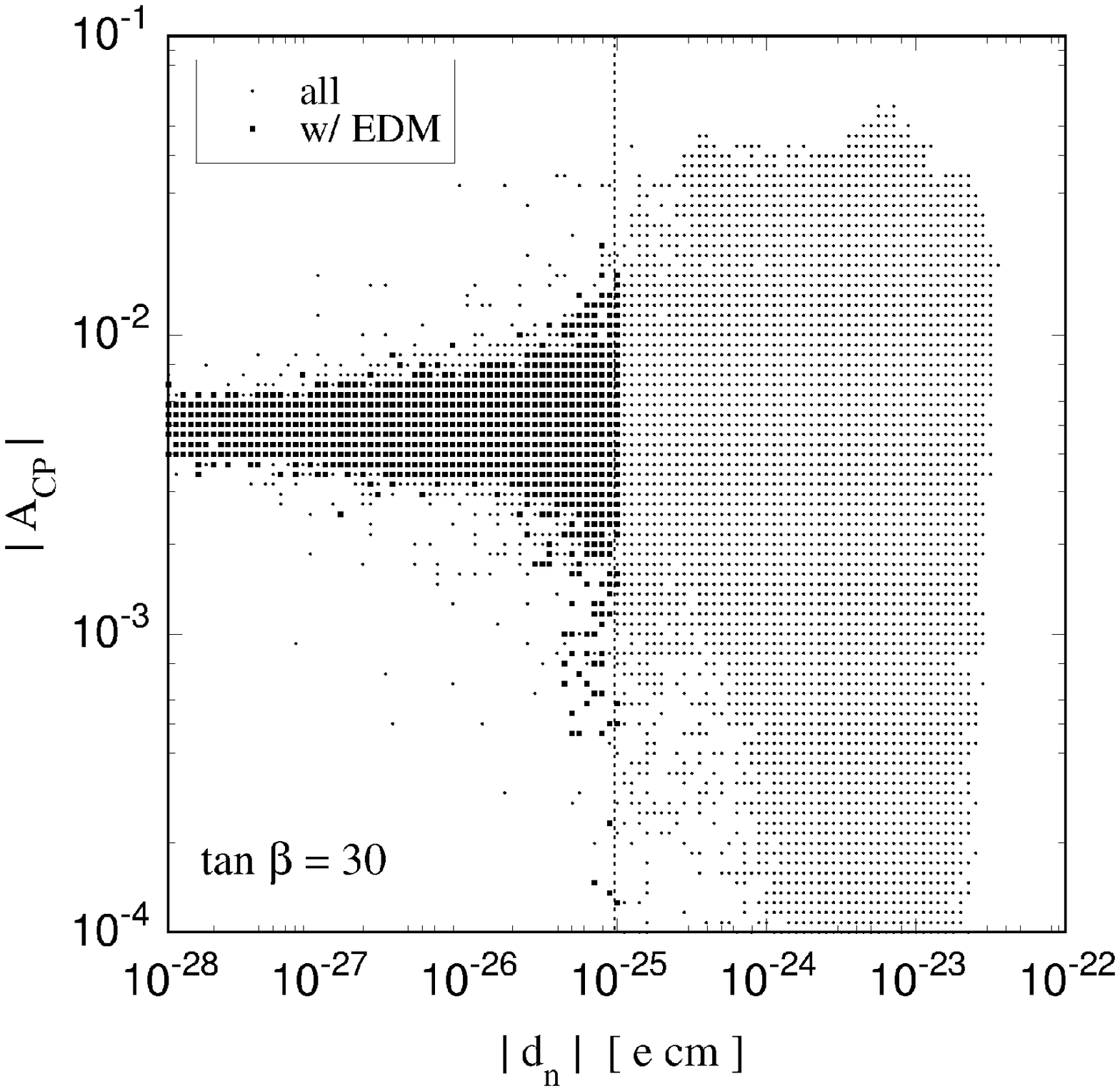}
}
\vfill
{\Large\bf Fig.~\ref{fig:bsllAcp-dn}(b)}
\end{center}
\clearpage

\end{document}